\documentclass[conference]{IEEEtran}

\usepackage{amsmath}
\usepackage{graphicx}
\usepackage{hyperref}
\usepackage{cite}
\usepackage{tikz}

\title{A Comparative Analysis of Machine Learning Models for Intrusion Detection in Intelligent Transport Systems}

\author{
\IEEEauthorblockN{Zawad Yalmie Sazid\textsuperscript{1}, Robert Abbas\textsuperscript{2}, Sasa Maric\textsuperscript{3}}
\IEEEauthorblockA{\textsuperscript{1,2}Victoria University, Sydney, Australia \\
\textsuperscript{3}University of New South Wales College, Sydney, Australia \\
Email: zawad.sazid@live.vu.edu.au, robert.abbas@vu.edu.au, s.maric@unsw.edu.au}
}

\begin{document}

\maketitle

\begin{abstract}
AI-powered edge computing security is moving Intelligent Transportation Systems (ITS) from passive, rule-based protections to proactive, smart, zero-touch, self-sufficient safeguards that neutralize threats in milliseconds. As transportation becomes more connected with edge computing, massive IoT, and advanced 5G for vehicle-to-everything (V2X) connectivity, AI at the edge computing nodes plays a crucial role in protecting against sophisticated threats, enabling URLLC (ultra-low-latency communications) for smart transport, and enhancing infrastructure capabilities and safety.

This research applies edge computing to improve latency, bandwidth efficiency, and service responsiveness by moving processing closer to devices, gateways, and users. However, this shift also expands the cyberattack surface because edge nodes are distributed, heterogeneous, and often resource-constrained. The paper proposes a trust-aware federated hybrid intrusion detection framework in which a random forest, a decision tree, and a linear SVM network learn complementary traffic representations at each edge site, while a server performs trust-aware aggregation of local model updates.
\end{abstract}

\begin{IEEEkeywords}
ITS, V2X, AI, zero-touch cybersecurity, 5G advanced-IoT, edge computing, URLLC, IDS, federated learning, hybrid deep learning, trust-aware aggregation, Random Forest, decision tree, and linear SVM.
\end{IEEEkeywords}

\section{Introduction}

Edge computing has been an effective answer to the latency, bandwidth, and privacy constraints of cloud-only architectures. Edge computing moves storage, analytics, and decision-making closer to the devices generating data. Instead of sending each packet, sensor reading, or control signal to a remote data center, edge computing aims to bring those functions nearer the harvester. This architectural transformation enables industrial automation, healthcare, and smart transport systems to support vast Internet of Things rollouts, but it also alters the cybersecurity challenge. Attack surfaces become more distributed, administrative boundaries multiply, and many edge devices operate with limited memory, energy, and processing capacity \cite{Shi2016,Satyanarayanan2017}.

Traditional signature-based intrusion detection systems remain useful, but by themselves, they are no longer sufficient for modern edge environments. They struggle with encrypted traffic, evolving attack behavior, heterogeneous endpoints, and distributed decision-making. Machine learning and deep learning methods therefore receive growing attention because they can learn statistical structure from traffic features and generalize across attack patterns that are difficult to encode manually. However, high-performing centralized models often assume raw-data aggregation, which conflicts with the privacy, bandwidth, and resilience requirements of edge environments \cite{Ferrag2022,Li2022}.

This paper proposes a trust-aware federated hybrid intrusion detection framework, while the presentation deck distilled the core problem, method, and preliminary achieved results. This revised final paper combines both sources into one cleaner academic document. It adds explicit research questions, relocates literature-comparison graphics to the literature review, places achieved result graphs in the experiment section, and rebuilds the key architecture figures so their labels are readable and their arrows remain non-overlapping.

The aim of this research is to design an intrusion detection framework for edge computing that can retain strong detection quality while preserving privacy, remaining suitable for distributed deployment, and reducing the influence of unreliable client updates. This aim leads to four practical objectives:
\begin{itemize}
    \item To design a hybrid local detector that combines denoised representation learning, local feature-pattern extraction, and cross-feature dependency modeling.
    \item To use federated learning so that edge sites collaborate without sharing raw traffic traces.
    \item To introduce trust-aware aggregation so that the server weights client updates by both data volume and reliability.
    \item To evaluate the framework using realistic intrusion-detection datasets and clearly interpret the achieved baseline results before moving to full federated implementation.
\end{itemize}

The research is guided by the following questions:
\begin{itemize}
    \item To what extent can federated learning preserve privacy in edge intrusion detection without requiring raw-data centralization?
    \item Can trust-aware aggregation reduce the influence of uneven, low-quality, or potentially malicious client updates compared with standard sample-size-weighted aggregation?
    \item What do the achieved baseline results on CICIDS2017 reveal about feature separability, model behavior, and the practical requirements for the next stage of federated evaluation?
\end{itemize}

\section{Literature Review}

Cybersecurity issues in distributed networks are a topic of extensive literature, which identifies the importance of fast, dependable detection schemes with minimal resource overhead. This section discusses the evolution of edge computing in Intelligent Transport Systems (ITS), known challenges for deploying intrusion detection systems (IDS) in resource-constrained environments, and why ensemble-based classical machine learning, specifically Random Forest, is a strong solution compared with baseline models such as Decision Tree and Linear SVM.

\subsection{Edge Computing and Security Challenges in ITS}

Early studies in edge computing proposed its need to overcome typical latency and bandwidth limitations of conventional centralized, cloud-centric architectures \cite{Satyanarayanan2017}. It also allows distribution of storage, analytics, and decision-making very close to the data source \cite{Shi2016}, which facilitates the real-time processing that modern applications require. In Intelligent Transport Systems environments, edge nodes such as Roadside Units (RSUs) and vehicle sensors enable ultra-real-time Vehicle-to-Everything (V2X) communications.

However, this architectural transition generates major security weaknesses. Edge nodes are more widely distributed and heterogeneous, and they do not usually have the same level of physical and network protection as centralized data centres \cite{Hozouri2025}. In an ITS environment, compromising an edge node does not only lead to data leakage \cite{Zhukabayeva2025}; it can also influence traffic routing and potentially harm passengers. As a result, the security model for ITS has to account for battered endpoints, constant connectivity issues, and a wide attack surface, meaning defences have to work locally and offline.

\subsection{Intrusion Detection Systems in Edge Networks}

Traditional signature-based IDS are increasingly insufficient for modern edge environments because they struggle with encrypted traffic, novel attack behaviours, and the diverse telemetry generated by IoT and vehicular sensors \cite{Rahman2025}. To address these shortcomings, the focus has shifted toward anomaly-based IDS powered by artificial intelligence \cite{Baidar2025}.

Despite the enthusiasm for AI-driven security, a fundamental operational challenge remains: edge devices possess limited memory, energy, and computational capacity. While highly complex models can achieve high theoretical accuracy, they impose significant processing overhead that induces unacceptable latency in real-time ITS networks. Therefore, researchers emphasize the operational value of deploying lightweight machine learning models that can function effectively within strict hardware constraints \cite{Mahadevappa2024}.

\subsection{Benchmark Datasets for Network Security}

An edge IDS cannot be evaluated credibly on outdated traffic. The research community has developed several benchmarks to test security models against contemporary threats. Datasets like UNSW-NB15 remain foundational because they encompass diverse, modern attack categories and avoid the pitfalls of legacy datasets \cite{Moustafa2015}. More recently, datasets such as Edge-IIoTset have been designed specifically to reflect the realities of IoT and Industrial IoT applications \cite{Ferrag2022}.

For evaluating the discriminative power of classical machine learning algorithms on enterprise and high-volume flow data, the CICIDS2017 dataset is heavily utilised. It contains thoroughly labelled benign traffic interspersed with common modern attacks, including DDoS, botnet, and brute-force campaigns \cite{Sharafaldin2018}. This makes it an ideal environment for testing whether a lightweight model can reliably separate malicious traffic from massive volumes of normal network flows.

\subsection{Evaluating Classical Machine Learning at the Edge}

Because of the processing limitations inherent to ITS edge nodes, classical machine learning models such as Random Forest, Decision Tree, and Support Vector Machines are often preferred over computationally prohibitive deep learning architectures. Studies have demonstrated that streamlined security models can achieve robust accuracy with minimal training time, proving the viability of simpler algorithms at the edge \cite{Mahadevappa2024}.

However, not all classical models perform equally under the stress of complex network traffic. Linear models such as Linear SVM are highly efficient but often struggle to capture the non-linear, high-dimensional relationships present in modern attack vectors, which frequently leads to an elevated rate of false positives. Conversely, single-estimator models like the Decision Tree provide excellent interpretability and fast inference speeds. Unfortunately, Decision Trees are prone to overfitting their training data, resulting in high variance and unstable performance when exposed to the dynamic, unpredictable traffic patterns of a vehicular network.

\subsection{The Superiority of Random Forest}

To resolve the limitations of standalone algorithms, ensemble methods have emerged as a strong choice for edge-based intrusion detection. Random Forest addresses the weaknesses of both Linear SVM and single Decision Trees by constructing a multitude of decision trees during the training phase and outputting the mode classification of the individual trees.

This ensemble approach systematically reduces variance and prevents overfitting, allowing the model to handle massive feature spaces and non-linear data without failing. In the context of an Intelligent Transport System, this translates to a model that maintains exceptionally high precision and recall while operating within the computational limits of an edge gateway. By maximizing attack capture while simultaneously suppressing false alarms, Random Forest minimizes analyst fatigue and improves reliability, making it a strong architectural choice for edge cybersecurity.

\section{Research Problem and Significance}

The research primarily considers the following problem: how can an Intelligent Transport Systems intrusion detection framework be deployed in a manner that enables resilient, high-accuracy attack identification without compromising vehicle privacy or introducing excessive computational weight that prevents deployment on resource-constrained edge nodes such as Roadside Units and connected vehicles? Fundamentally, the problem is that contemporary ITS scenarios take advantage of instantaneous Vehicle-to-Everything communications to regulate the flow of traffic, reduce accidents, and provide functionality for autonomous vehicle operation. This reduces latency and bandwidth bottlenecks by moving analytics and decision-making to the edge. However, this distributed structure also increases the attack surface.

Existing centralized detection systems are sub-optimal because they breach user privacy by transferring raw vehicular telemetry to the cloud and also consume significant bandwidth. Federated learning is privacy-friendly because it helps keep the data local, but selecting the correct AI-based model for edge nodes continues to be a major challenge. Deep learning models are often too computationally intensive for edge hardware.

Conversely, while classical machine learning shows promise for edge deployments, models like Linear SVM struggle to map the complex, non-linear traffic patterns of modern cyberattacks, and single Decision Trees are prone to overfitting and generating high false-alarm rates.

The significance of solving this problem is substantial. In an Intelligent Transport System, a cyberattack transcends data loss; it presents an immediate physical safety hazard. Compromised edge controllers or flooded vehicle networks, such as in DDoS attacks, can lead to unsafe physical control actions, traffic gridlock, or severe accidents. Therefore, a practical ITS security model must provide immediate, highly reliable discrimination between benign and malicious traffic. It requires a model that minimizes false positives to prevent unnecessary vehicle braking or alert fatigue, all while operating efficiently within a federated edge architecture.

This study contributes to the field of edge-secured Intelligent Transport Systems in three primary ways. First, by defining ITS edge constraints, the research systematically evaluates the operational viability of classical machine learning algorithms for federated edge-based intrusion detection, intentionally pivoting away from computationally prohibitive deep learning models to ensure low-latency performance on RSUs and vehicle onboard units.

Second, it identifies a promising AI-powered solution by demonstrating that Random Forest, an ensemble of decision trees, successfully resolves both the high-variance behaviour typical of single Decision Trees and the classification limitations that cause Linear SVMs to struggle in complex network traffic scenarios.

Finally, through empirical error profiling grounded in modern, realistic datasets like CICIDS2017, the study provides a detailed analysis showing that Random Forest not only achieves near-perfect accuracy but also dramatically reduces the operational burden of false alarms. This clean error profile establishes it as a strong foundation for mission-critical, federated intrusion detection in Intelligent Transport Systems.

\begin{figure*}[]
    \centering
    \includegraphics[width=\textwidth]{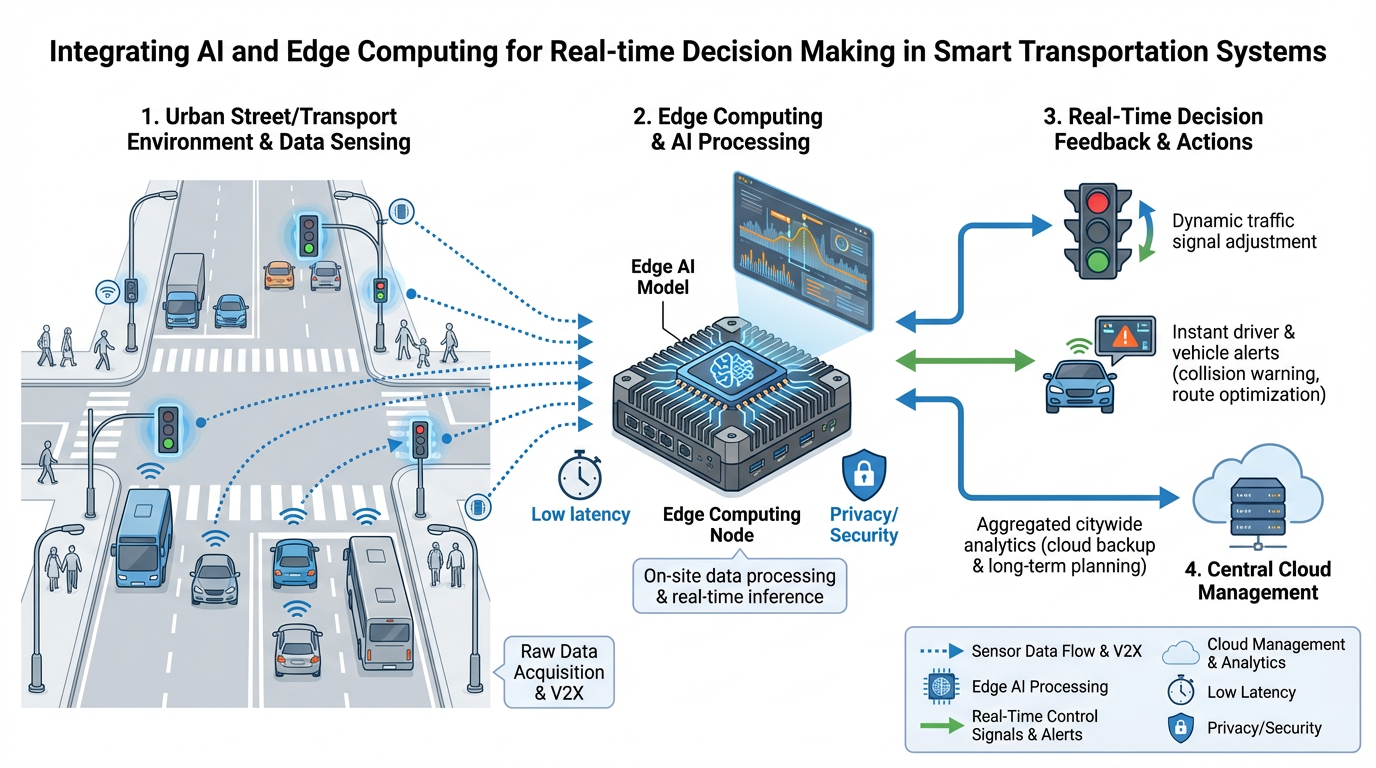}
    \label{fig:ai_edge}
\end{figure*}

\section{System Model and Methodology}

Edge computing aims to bring computing resources closer to where data is generated. It is ideally suited for circumstances requiring low latency or real-time processing, as well as situations in which massive amounts of data are transported to a central point without necessity.

To meet the networking and processing demands of a future world with connected 5G-IoT and 6G-IoT, edge computing is about bringing intelligence and control closer to endpoints. Figure~\ref{fig:ai_edge} presents an integration of AI and edge computing for zero-touch security and real-time decision-making for ITS.

Edge computing is a distributed framework in which compute capabilities such as processing, analysis, and storage are moved to the edge of a network, geographically closer to where the data is being generated or consumed. This results in less data being transmitted between devices and centralized data centres, reduced network congestion, and shorter data transit distances. As a result, edge computing offers advantages such as high bandwidth, low latency, and greater control over data management and sovereignty.

\subsection{Overall Architecture}

The proposed framework is a trust-aware federated intrusion-detection system for edge environments. Each edge site captures and preprocesses its own traffic, trains a local hybrid detector, and sends only model updates and trust-related validation indicators to a coordinating server. The server evaluates trust, aggregates local updates, and redistributes the updated global model. Raw traffic never leaves the site. This design directly supports environments where privacy, bandwidth, and availability make raw-data centralization undesirable.

\begin{figure*}[t]
    \centering
    \includegraphics[width=\linewidth]{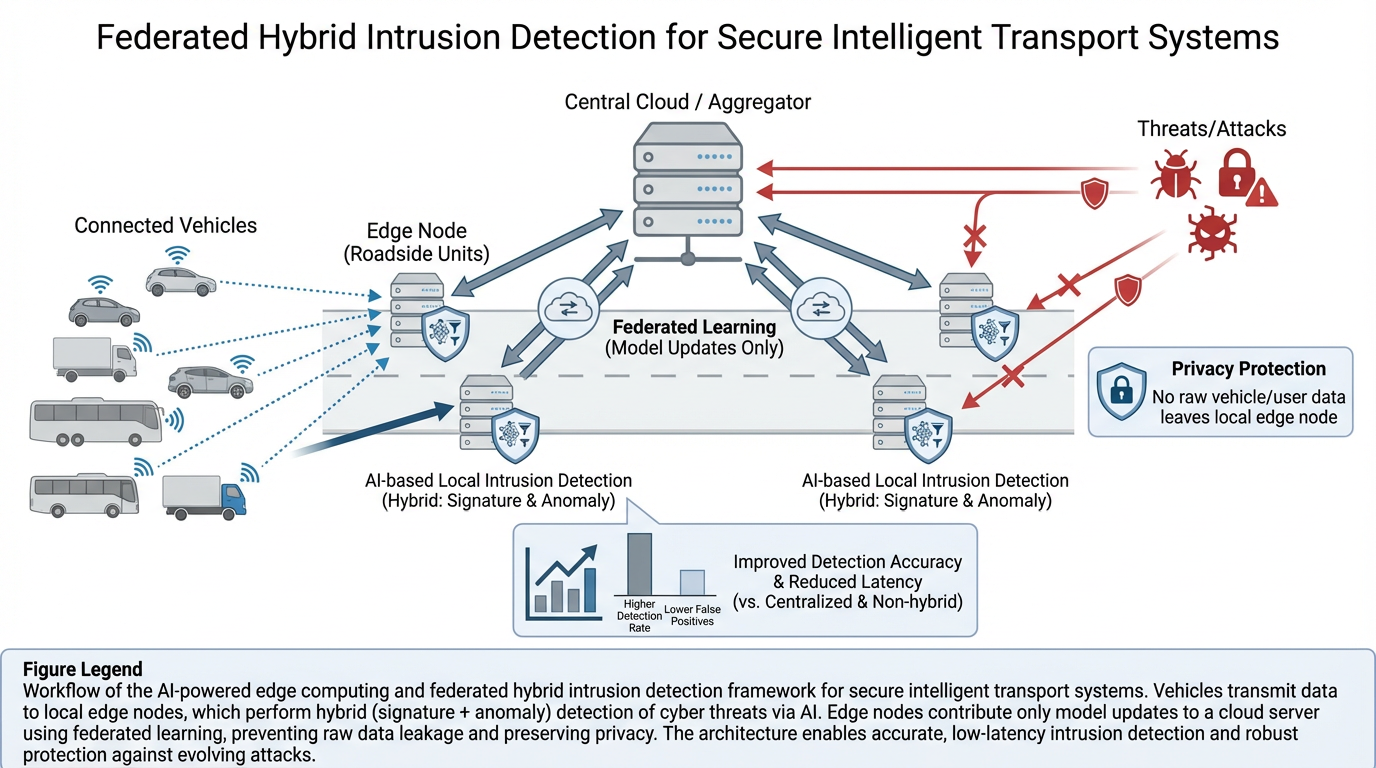}
    \caption{IDS-powered EC-FL V2X.}
    \label{fig:v2x_ec_fl}
\end{figure*}

\begin{figure*}[t]
    \centering
    \includegraphics[width=\linewidth]{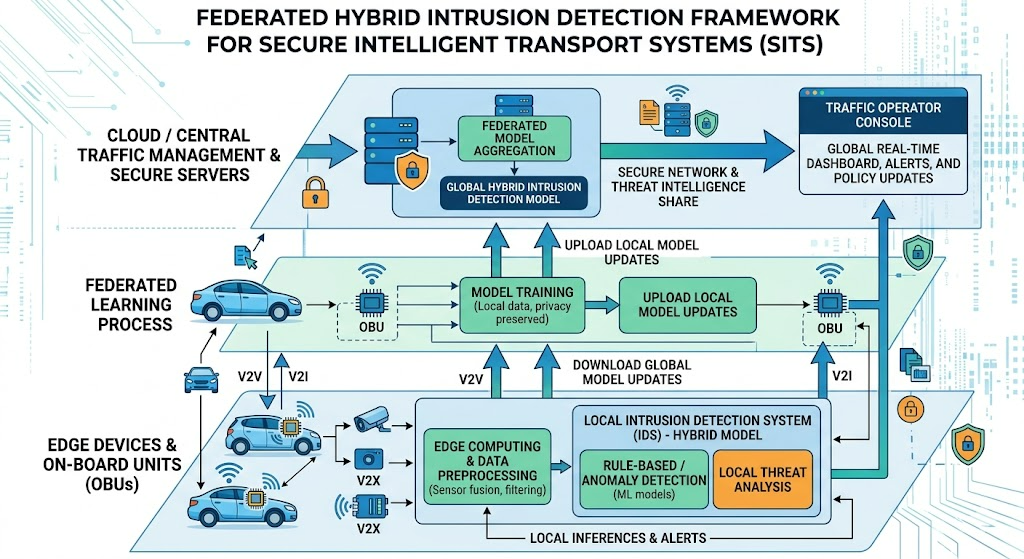}
    \caption{SITS.}
    \label{fig:sits}
\end{figure*}

\textbf{Operational workflow:} The end-to-end workflow of the framework is deliberately simple. Traffic is captured and transformed into a common feature representation, local hybrid models are trained privately, model updates are transmitted to the server, trust-aware aggregation is applied, and the updated global model is returned to the clients. This loop repeats over federated rounds until convergence.

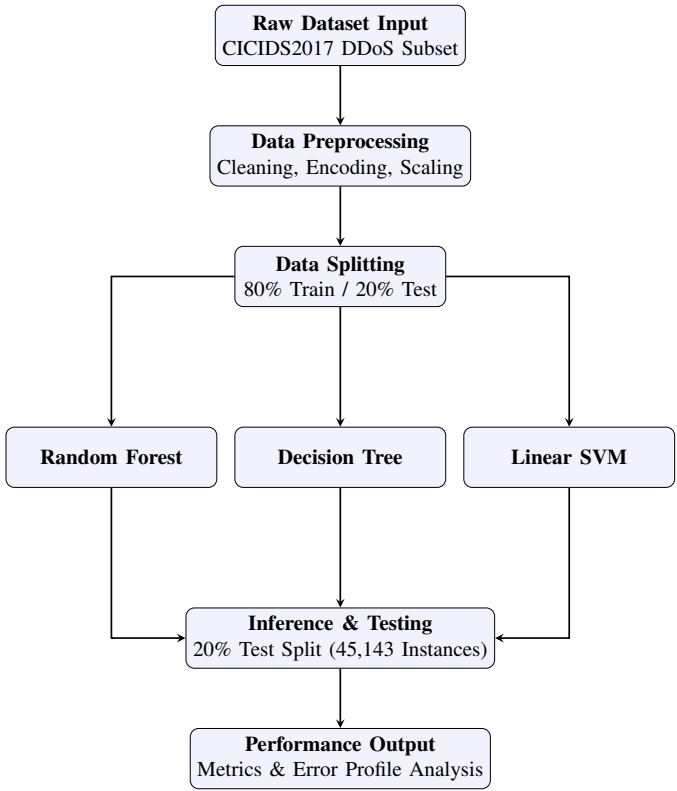
\begin{figure}[htbp]
\centering
\resizebox{\linewidth}{!}{%
\begin{tikzpicture}[node distance=2cm]
\tikzstyle{block} = [rectangle, rounded corners, minimum width=3.5cm, minimum height=1cm, text centered, draw=black, fill=blue!5, align=center]
\tikzstyle{arrow} = [thick,->,>=stealth]

\node (data) [block] {\textbf{Raw Dataset Input} \\ CICIDS2017 DDoS Subset};
\node (prep) [block, below of=data] {\textbf{Data Preprocessing} \\ Cleaning, Encoding, Scaling};
\node (split) [block, below of=prep] {\textbf{Data Splitting} \\ 80\% Train / 20\% Test};

\node (dt) [block, below of=split, yshift=-1cm] {\textbf{Decision Tree}};
\node (rf) [block, left of=dt, node distance=3.8cm] {\textbf{Random Forest}};
\node (svm) [block, right of=dt, node distance=3.8cm] {\textbf{Linear SVM}};

\node (test) [block, below of=dt, yshift=-1cm] {\textbf{Inference \& Testing} \\ 20\% Test Split (45,143 Instances)};
\node (output) [block, below of=test] {\textbf{Performance Output} \\ Metrics \& Error Profile Analysis};

\draw [arrow] (data) -- (prep);
\draw [arrow] (prep) -- (split);
\draw [arrow] (split) -- (dt);
\draw [arrow] (split) -| (rf);
\draw [arrow] (split) -| (svm);
\draw [arrow] (rf) |- (test);
\draw [arrow] (dt) -- (test);
\draw [arrow] (svm) |- (test);
\draw [arrow] (test) -- (output);
\end{tikzpicture}%
}
\caption{Workflow diagram of the intrusion detection machine learning pipeline.}
\label{fig:ml_workflow}
\end{figure}

\section{Data Collection and Processing}

This research used the CICIDS2017 dataset, which contains flow-based enterprise traffic and common modern attacks \cite{Sharafaldin2018}. The quantitative baseline available in this paper comes from CICIDS2017.

The CICIDS2017 dataset is a large, labelled network traffic dataset created by the Canadian Institute for Cybersecurity in 2017 to evaluate intrusion detection systems. It contains modern benign and malicious traffic, including 14 attack types such as DoS, DDoS, brute force, and botnet behaviour, developed over five days and totalling more than 2.8 million records.

\textbf{Key characteristics:}
\begin{itemize}
    \item \textbf{Data structure:} Labelled flows in CSV format extracted using CICFlowMeter and also available as PCAP files.
    \item \textbf{Traffic types:} Fourteen attack categories plus benign traffic.
    \item \textbf{Features:} Includes approximately 77--83 features, including source and destination IP, ports, protocols, and bidirectional flow statistics.
    \item \textbf{Size:} More than 2.8 million instances.
    \item \textbf{Protocols:} Includes email, SSH, FTP, HTTP, and HTTPS traffic.
\end{itemize}

\textbf{Typical attacks included:}
\begin{itemize}
    \item Brute force: SSH-BruteForce and FTP-BruteForce.
    \item DoS/DDoS: DoS Slowhttptest, DoS GoldenEye, DoS Hulk, and DoS Slowloris.
    \item Web attacks: Infiltration, Heartbleed, Botnet, Web Attack-Brute Force, Web Attack-XSS, and Web Attack-SQL Injection.
\end{itemize}

This dataset is commonly used to train machine learning and deep learning models to discriminate between benign and malicious network activity. While useful, the original dataset frequently requires preprocessing to correct missing values, duplicate entries, and class imbalance.

\subsection{Evaluation Metrics}

This study employs five standard classification metrics to comprehensively assess model performance across different operational priorities. Each metric captures distinct aspects of classifier behavior essential for security applications [22]:
\begin{enumerate}
    \item \textbf{Accuracy:} Measures the percentage of total predictions that are correctly classified by the model:
    \begin{equation}
    \frac{TP+TN}{TP+TN+FP+FN}
    \end{equation}

    \item \textbf{Precision:} Measures the reliability of attack predictions by assessing how many flagged alerts are actually correct:
    \begin{equation}
    \frac{TP}{TP+FP}
    \end{equation}

    \item \textbf{Recall:} Measures the model's ability to identify true attack instances, reflecting the rate of missed detections:
    \begin{equation}
    \frac{TP}{TP+FN}
    \end{equation}

    \item \textbf{F1-score:} Combines precision and recall into a unified measure to reflect overall classification performance:
    \begin{equation}
    \frac{2 \times (\text{Precision} \times \text{Recall})}{\text{Precision}+\text{Recall}}
    \end{equation}

    \item \textbf{ROC-AUC:} Represents the area under the ROC curve, which graphs the true positive rate against the false positive rate across classification thresholds. It provides a threshold-independent assessment of discrimination power and estimates the model's capability to distinguish between classes at all operating points. A ROC-AUC score varies from 1.0 for complete separation to 0.5 for random guessing. Scores above 0.9 represent excellent discrimination.
\end{enumerate}

ROC-AUC is particularly well-suited for IDS assessment because it provides performance evaluation across operating points rather than at a single threshold. This threshold-independent capability enables analysts to select operating points that balance detection and false alarm rates depending on the attack surface.

\section{Results and Evaluation}

\subsection{Achieved Quantitative Performance}

These baseline results demonstrate that the chosen CICIDS2017 binary task is well separated in the handcrafted feature space. The best result was achieved with Random Forest, followed by Decision Tree and Linear SVM. The metrics achieved are summarized in Table~\ref{tab:performance_metrics}.

\begin{table*}[htbp]
    \centering
    \caption{Performance metrics}
    \label{tab:performance_metrics}
    \begin{tabular}{|c|c|c|c|c|c|}\hline
         Model & Accuracy & Precision & Recall & F1-score & ROC-AUC \\ \hline
         Random Forest & 0.999889 & 0.999922 & 0.999883 & 0.999902 & 1.00000 \\ \hline
         Decision Tree & 0.999756 & 0.999727 & 0.999844 & 0.999785 & 0.999833 \\ \hline
         Linear SVM & 0.999092 & 0.999258 & 0.999141 & 0.999199 & 0.999882 \\ \hline
    \end{tabular}
\end{table*}

\begin{figure}[htbp]
    \centering
    \includegraphics[width=\linewidth]{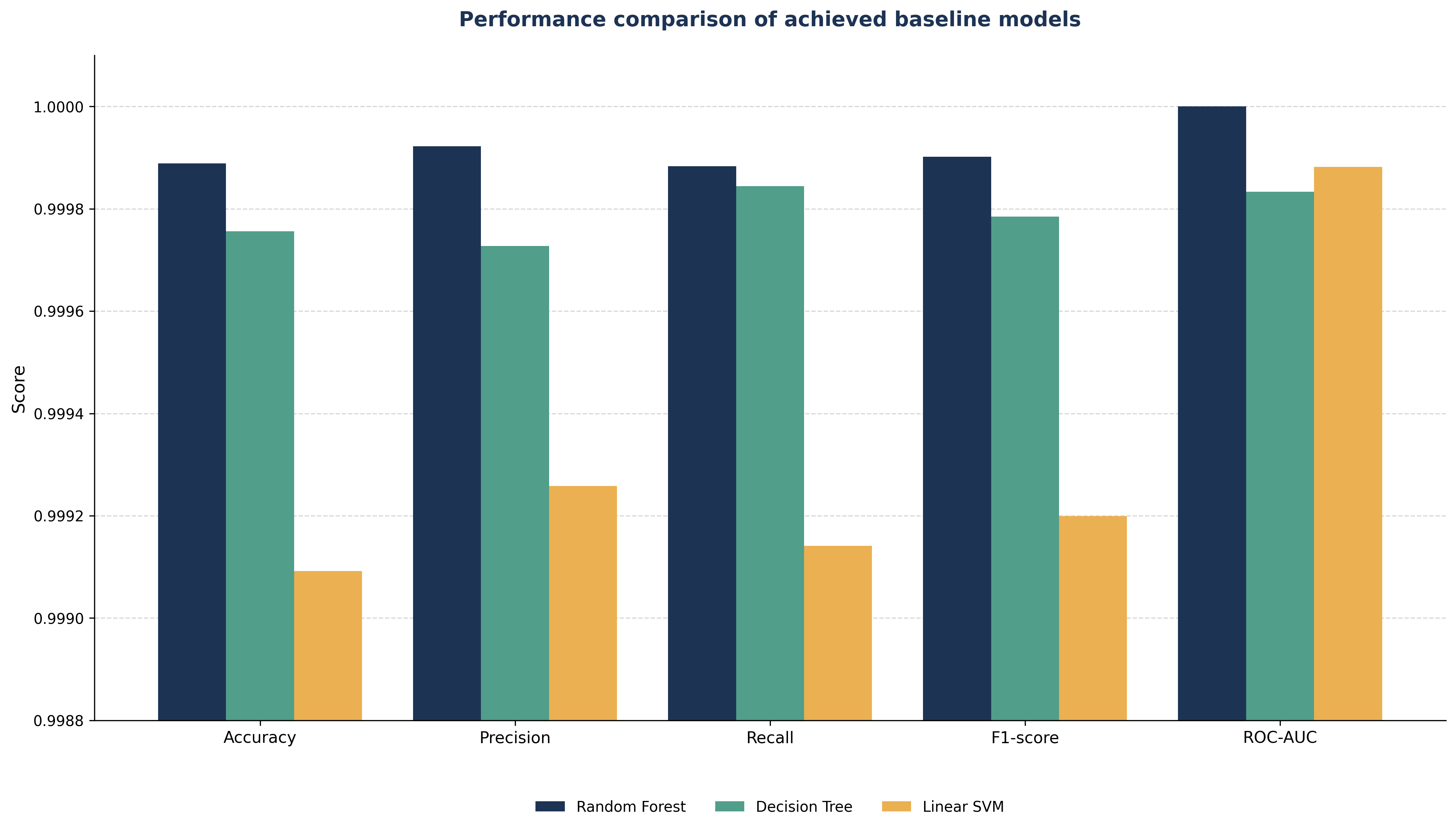}
    \caption{Achieved performance comparison across accuracy, precision, recall, F1-score, and ROC-AUC for Random Forest, Decision Tree, and Linear SVM.}
    \label{fig:model_comparison}
\end{figure}

Two things are clear from Figure~\ref{fig:model_comparison}. First, all three models achieve excellent prediction performance; this indicates that the selected feature representation contains effective discriminative information for differentiating benign flows from DDoS traffic. Second, Random Forest is the most balanced model across all five metrics. Its precision is the highest, so when it flags an attack, it is slightly more likely than the others to be correct. It also has a very high recall, so it misses very few attacks. By combining these two properties, it attains the highest F1-score.

Decision Tree also performed well and can still be a strong baseline for deployments with limited resources. It has a recall slightly higher than its precision, indicating that it is effective at capturing attacks but generates more false positives than Random Forest. Although Linear SVM still scores very high overall, its metric values are consistently lower than the other two models. This difference is small in absolute terms but operationally significant, because even a small increase in false positives or missed detections can increase incident response cost at scale.

\subsection{Class Support and Result Interpretation}

\begin{figure}[htbp]
    \centering
    \includegraphics[width=\linewidth]{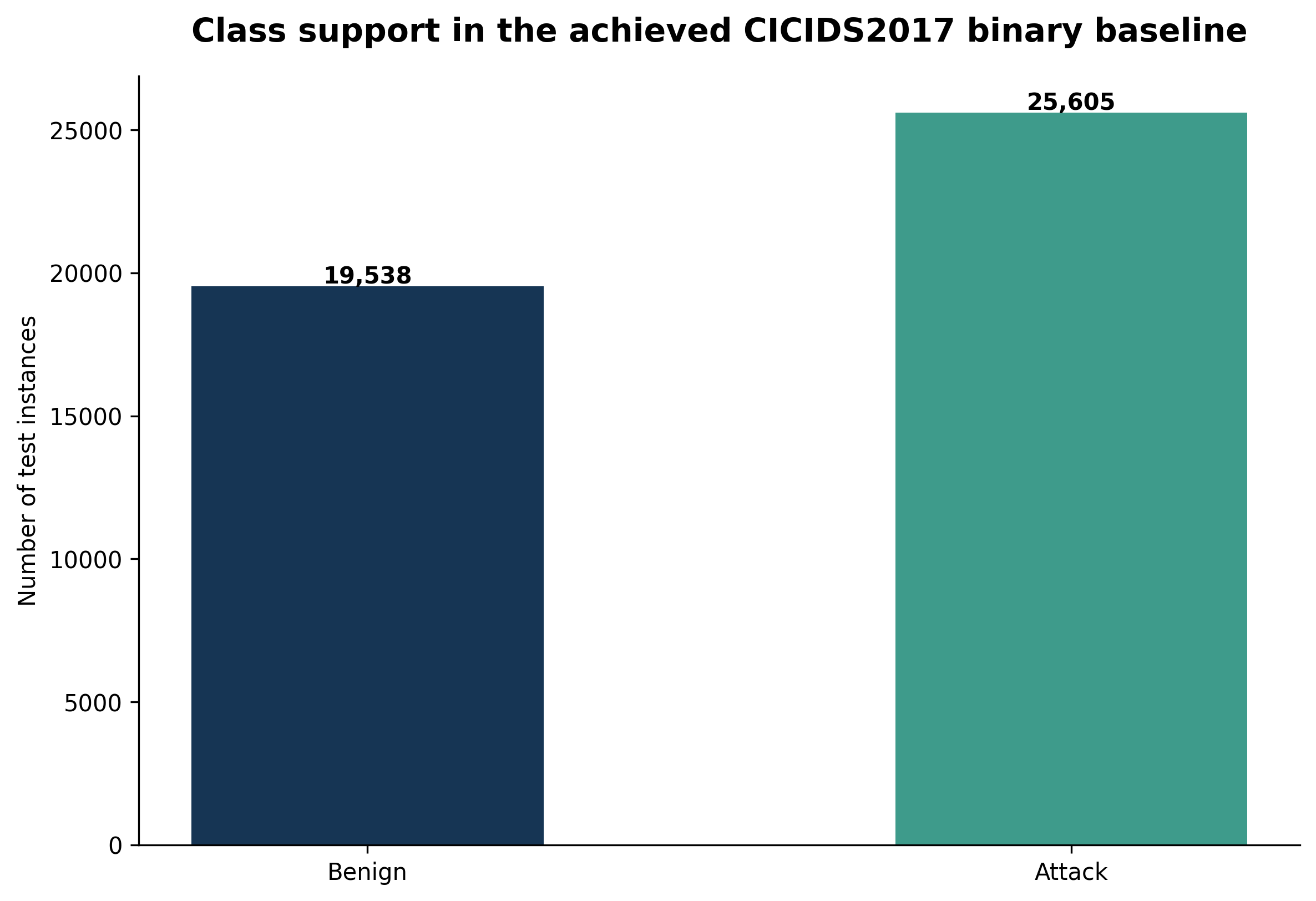}
    \caption{Class support in the achieved CICIDS2017 binary baseline.}
    \label{fig:class_support}
\end{figure}

As seen in Figure~\ref{fig:class_support}, the achieved test set consists of 19,538 benign instances and 25,605 attack instances. This is not perfectly balanced, but the binary task has only slight skew. This is important because the near-perfect result is not simply a function of overwhelming majority-class overrepresentation. Instead, the results indicate that these two classes are well separated in the selected feature space.

That interpretation matters for the larger research agenda. Strong local separability means the next phase of exploration need not be whether the traffic is learnable at all, but whether privacy-preserving distributed learning can sustain that performance under realistic edge constraints. Using strong baseline results reduces uncertainty around feature quality, so further research can focus on federation, trust weighting, and deployment trade-offs.

\subsection{Confusion Matrix and Error-Profile Discussion}

\begin{figure}[htbp]
    \centering
    \includegraphics[width=\linewidth]{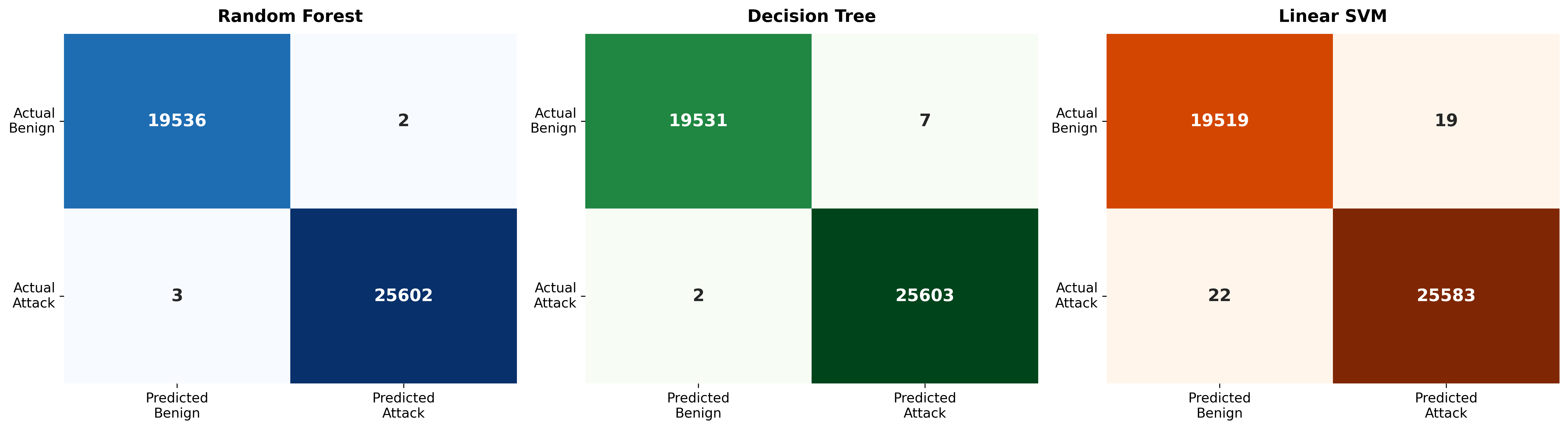}
    \caption{Confusion matrices for the three achieved baseline models.}
    \label{fig:confusion_matrices}
\end{figure}

\begin{figure}[htbp]
    \centering
    \includegraphics[width=\linewidth]{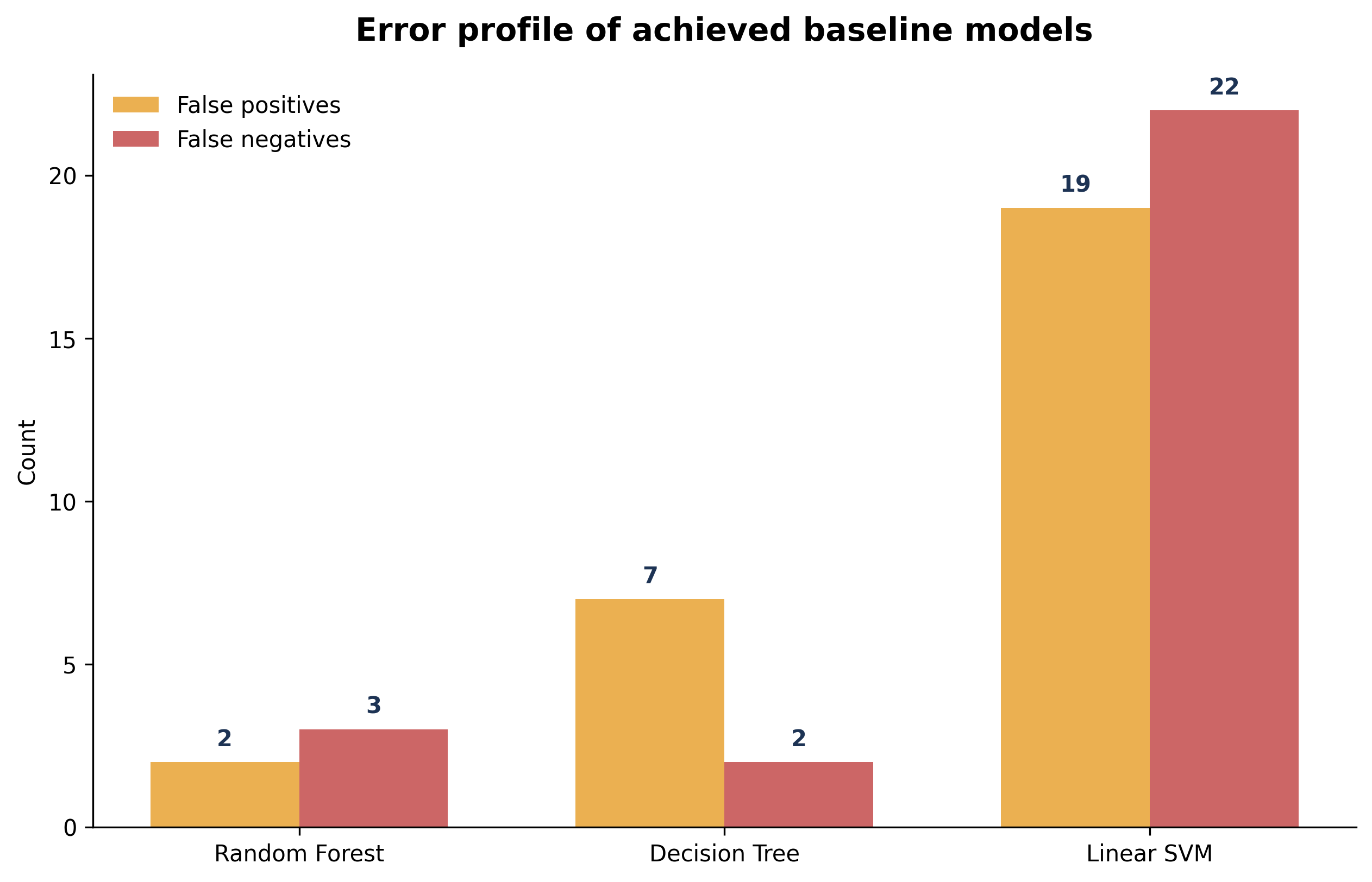}
    \caption{False-positive and false-negative profiles derived from the achieved confusion matrices.}
    \label{fig:error_profile}
\end{figure}

The confusion matrices in Figure~\ref{fig:confusion_matrices} provide a more operationally meaningful view of model behaviour than aggregate metrics alone. Random Forest has 19,536 true negatives and 25,602 true positives, but only 2 false positives and 3 false negatives. This is an impressively clean error profile and explains why it tops the metric table. In the context of a cybersecurity deployment, low false positives help decrease analyst fatigue and low false negatives help ensure that real attacks are not missed.

Decision Tree remains very competitive. It produces 7 false positives and 4 false negatives, which is still a very good result but not as clean as Random Forest. The slightly increased number of false positives means it would still generate benign alerts that need to be reviewed. In some edge contexts, that may be acceptable if computational simplicity is prioritised, but it still represents a practical trade-off.

Linear SVM produces 19 false positives and 22 false negatives, which is the largest error count among the achieved baselines. Although its overall accuracy remains very high, Figure~\ref{fig:error_profile} shows that its operational error burden is clearly higher than that of the tree-based models. This is why a detailed confusion-matrix discussion is necessary: a model can appear excellent in headline accuracy while still imposing more review cost and greater missed-attack risk than an alternative model.

\subsection{What the Achieved Results Mean for the Proposed Framework}

\begin{figure}[htbp]
    \centering
    \includegraphics[width=\linewidth]{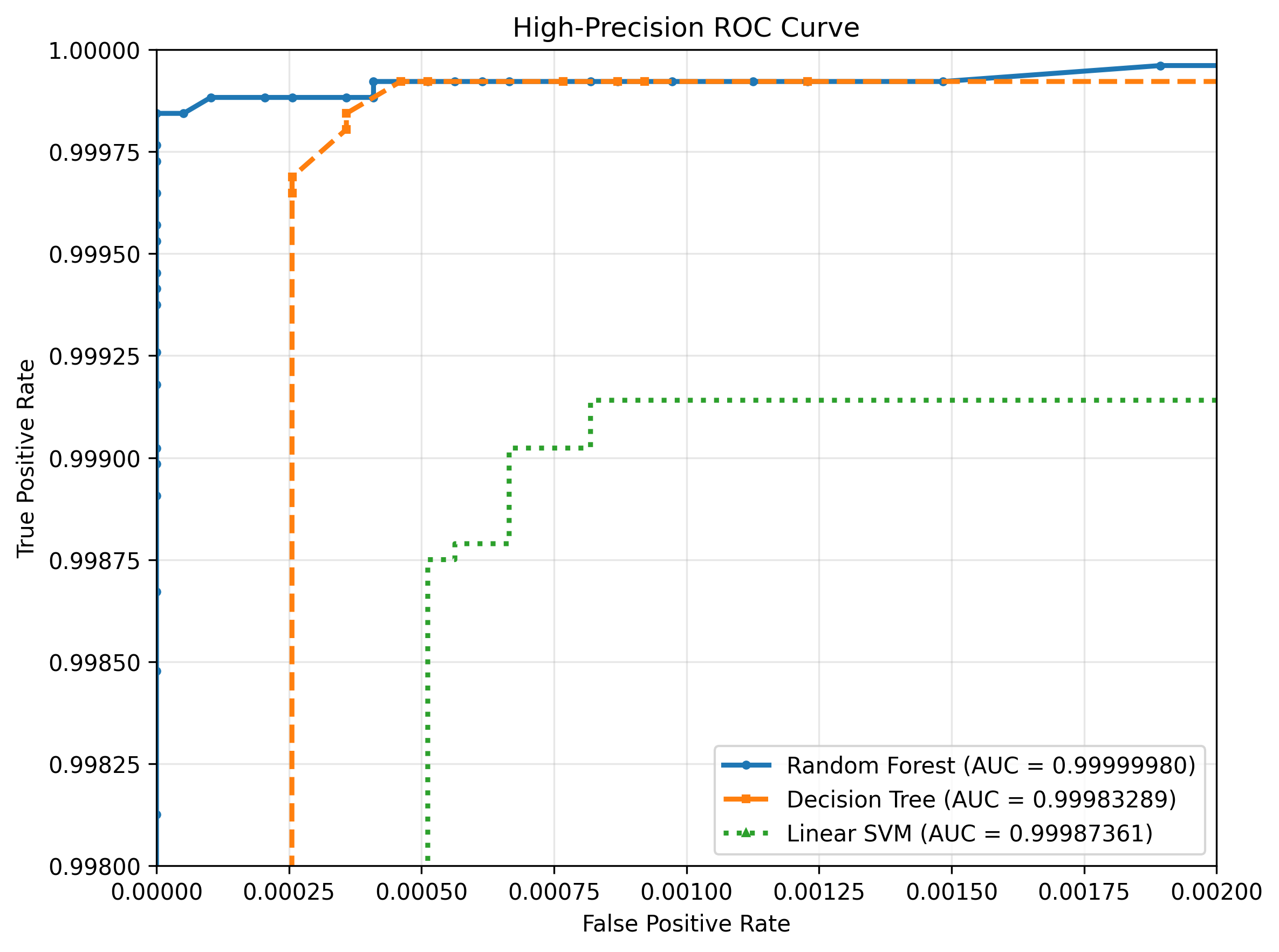}
    \caption{ROC curve comparison for Random Forest, Decision Tree, and Linear SVM on the CICIDS2017 DDoS binary classification task.}
    \label{fig:roc_curve}
\end{figure}

The ROC curves in Figure~\ref{fig:roc_curve} show that all three baseline models achieved excellent discrimination between benign and attack traffic, with their curves concentrated near the upper-left corner of the graph. This indicates high true positive rates at very low false positive rates, which is ideal behaviour for an intrusion detection model. Random Forest performed best out of the three models with an ROC-AUC of 1.0000, followed by Linear SVM at 0.999882 and Decision Tree at 0.999833.

These values indicate that the chosen feature space is extremely effective for separating DDoS traffic from benign flows within the CICIDS2017 binary setting. One important observation from the figure is that the three ROC curves are very close to each other. This does not indicate a plotting problem. Instead, it shows that all three classifiers achieved near-perfect performance on this dataset split. Even so, Random Forest still has the best overall ranking based on ROC-AUC and the best balance across accuracy, precision, recall, and F1-score. This makes Random Forest the strongest classical baseline in this study.

From a research perspective, this result is important because it validates the discriminative strength of the engineered traffic features before moving to the proposed federated hybrid model. In other words, the baseline experiment shows that attack and benign classes are already well separated locally. The next research step is therefore not just to improve raw classification performance, but to test whether similarly strong detection can be maintained under federated, privacy-preserving, and non-IID edge conditions, where communication cost, client heterogeneity, and trust-aware aggregation become more important.

The results also support several findings directly relevant to the proposed trust-aware federated hybrid framework. First, the selected feature space is strong enough to justify moving beyond a single-site baseline. Second, Random Forest provides a demanding reference point for any future hybrid model: a more complex model should not be adopted merely because it is technically richer. It should add value through privacy preservation, cross-site generalization, or robustness under heterogeneity and unreliable clients.

Third, the results highlight the difference between local success and deployable edge security. The current evidence comes from a local binary baseline on one dataset subset. It does not yet answer how performance changes when data are non-IID across clients, when communication costs are constrained, or when some participants behave unreliably. That gap is precisely why the trust-aware federated architecture remains important despite the strong achieved baseline.

Finally, the achieved baseline clarifies the next implementation priorities. The full system should compare centralized hybrid learning, standard federated averaging, and trust-aware federated aggregation across multiple datasets. If the trust-aware federated model can preserve most of the local detection quality while improving privacy and controlling weak updates, then it will offer a stronger overall solution for edge cybersecurity than any isolated local baseline.

\section{Conclusion}

This research mainly aimed to identify an optimal and time-efficient machine learning classifier to protect edge nodes of Intelligent Transport Systems. This study used a comparative analysis of classical machine learning algorithms to show that Random Forest is the best-performing algorithm for intrusion detection at the edge when compared directly with Decision Tree and Linear SVM.

The experimental results verify that Random Forest is consistently better across the main metrics. Random Forest achieved an accuracy of 99.9889\%, precision of 99.9922\%, recall of 99.9883\%, F1-score of 99.9902\%, and ROC-AUC equal to 1.0000 on the binary classification task. These results outperformed the other models under study. The Decision Tree classifier achieved an accuracy of 99.9756\%, while Linear SVM achieved 99.9092\%.

When moving beyond headline metrics, the error profile clearly shows why Random Forest is superior operationally. Across more than 45,000 test instances, Random Forest generated only 2 false positives and 3 false negatives. By contrast, Decision Tree produced 7 false positives and 4 false negatives, whereas Linear SVM produced 19 false positives and 22 false negatives. Low false negatives are required to protect passenger and network integrity, while very low false positives help prevent over-alerting and unnecessary intervention in a real transportation network.

In the final analysis, these results show that the high-performance and low-latency cybersecurity required by modern Intelligent Transport Systems is within practical reach of the Random Forest algorithm. It strikes an effective combination of classification accuracy, stability, and low operational error that makes it the strongest overall baseline model for defending vehicular edge networks from malicious traffic.

\bibliographystyle{IEEEtran}

\end{document}